# TOT Measurement Implemented in FPGA TDC[*]

FAN Huan-Huan(范欢欢)[1,2; 1]    CAO Ping(曹平)[1,2; 2]
LIU Shu-Bin(刘树彬)[1,2]    AN Qi(安琪)[1,2]

[1] State Key Laboratory of Particle Detection and Electronics, University of Science and Technology of China, Hefei, 230026, China
[2] Anhui Key Laboratory of Physical Electronics, Department of Modern Physics, University of Science and Technology of China, Hefei, 230026, China

**Abstract:** Time measurement plays a crucial rule for the purpose of particle identification in high energy physical experiments. With the upgrading of physical goal and the developing of electronics, modern time measurement system meets the requirement of excellent resolution specification as well as high integrity. Due to Field Programmable Gate Array (FPGA), FPGA time-to-digital converter (TDC) becomes one of mature and prominent time measurement methods in recent years.  For correcting time-walk effect caused by leading timing, time-over-threshold (TOT) measurement should be added in the FPGA TDC. TOT can be obtained by measuring the interval time of signal leading and trailing edge. Unfortunately, a traditional TDC can recognize only one kind of signal edge, the leading or the trailing. Generally, to measure the interval, two TDC channels can be used at the same time, one for leading, the other for trailing. However, this method will increase the amount of used FPGA resource and reduce the TDC's integrity unavoidably.

This paper presents one method of TOT measurement implemented in a Xilinx Virtex-5 FPGA. In this method, TOT measure can be achieved in only one TDC input channel. The consumed resources and time resolution can both be guaranteed. Test shows that this TDC can achieve resolution better than 15ps for leading edge measurement and 37ps for TOT measurement. Furthermore, the TDC measuring dead time is about 2 clock cycles, which makes it be good for applications of higher physical event rate.

**Keywords:** time measurement, time-over-threshold (TOT) measurement, field programmable gate arrays (FPGA), time-to-digital converter (TDC)

**PACS:** 29. 85. Ca

## 1 Introduction[1]

Time measurement plays a crucial role for particle identification in high energy physical experiments, especially in the time-of-flight (TOF) detector [1], [2], [3]. Up to the present, two main kinds of technologies are developed to implement time measurement. One is the dedicated TDC ASIC, such as HPTDC designed by CERN [4]. The other is FPGA TDC [5] - [9].  Compared with ASIC TDC, FPGA TDC has advantages of shorter development time, higher flexibility and lower cost. It makes the FPGA TDC fully fledged in recent years.

Various kinds of timing methodology have been widely developed, such as leading edge timing (LET), constant fraction timing (CFT) etc. [10], [11], [12]. Comparing with CFT, the structure of LET is very simple. It only consists of a comparator and several passive elements. Due to its simplicity and easy to setup, LET becomes one of the most popular technologies to measure time. But the biggest drawback of this technology is that it has very large time walks. A timing mark which indicates the time of an occurred event is generated by LET as soon as the signal crosses the threshold level, but the timing mark inevitably shows a large dependency on the amplitude of the signal. This time walk brings in the dominant time measurement errors and deteriorates the time resolution of time measurement. Fortunately this unwanted influence can be corrected by the amplitude information of the signal. Because the charge amounts of the signal have mainly been used as amplitude information now, most applications perform time-walk corrections by measuring charge amounts instead of the amplitude.

The traditional technology is to apply two independent circuits to measure the signal's time and charge amounts. Dedicated TDC ASIC is preferred used to measure time for simplifying circuit. But charge amounts measurement circuit can only be constructed by the discrete components which make the back-end circuit very complex.

Especially high resolution and high density is the tendency of the high energy physical experiment in the future. For example, the electronic channel number of End-cap time-of-flight (ETOF) of the BESⅢ increases to 1728 after upgrade. It becomes almost twenty times of the primary 96 channels. In this case, it becomes very fatal and urgent to simplify the electronic system and increase its integrity. This not only makes it easy to setup the whole system, but also reduces costs.

Now it emerges many dedicated leading edge discriminators (LEDs) for time measurement used in the time-of-flight detector, which is based on a time-over-threshold approach such as NINO ASIC [13], [14]. The outputs of these discriminators reflect the event's time information in the rising edge and the charge amounts in the pulse width. If one device has the capability of time measurement and pulse width measurement, it will simplify the whole TDC measurement circuit greatly combining with LEDs.

---

[1]  * Supported by National Natural Science Foundation of China (11079003,10979003)
1) E-mail: fanhh@mail.ustc.edu.cn
2) E-mail: cping@ustc.edu.cn

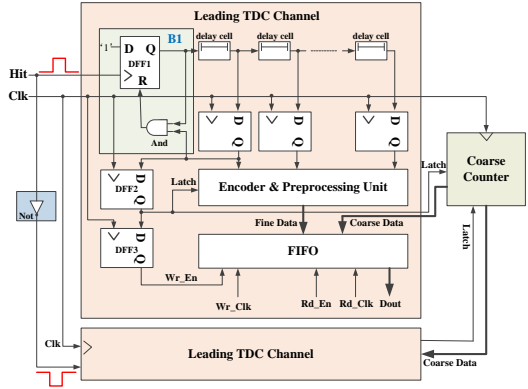

Fig. 1 The block diagram of pulse width measurement with two TDC channels

In our previous work [8], a high resolution TDC has been implemented in Xilinx FPGA. It applies the methodology of combining coarse time measurement with fine time measurement. Coarse time measurement utilizes normal binary counter to measure coarse time which is a multiple of the clock period. The fine time which is less than one clock period is measured by fine time measurement. It utilizes dedicated carry chain resource in Xilinx FPGA to construct delay chain for time interpolation and achieves time measurement with precision better than 15ps. It also achieves the pulse width measurement with the resolution of about 27ps through reversing the input signal and measuring its trailing edge in another TDC channel which is shown in Fig.1. But this increases the same amount of resource to measure time and pulse width which reduces the TDC's integrity and increases the cost.

In this paper, we describe a high resolution and high density method of TOT measurement implemented on our evaluation board with a XC5VLX220FF1760 FPGA from Xilinx Virtex-5 family. It doesn't need extra resource and one TDC channel is enough to measure both time and pulse width.

The structure and implementation of the TOT measurement is presented in Section Ⅱ. The test result will be briefly introduced in Section Ⅲ. Finally Section Ⅳ gives a conclusion.

## 2 Implementation of TOT Measurement

Base on we have implemented time measurement in Xilinx Virtex-5 FPGA, the simplest thought for measuring pulse width is to respectively obtain leading edge time and trailing edge time, then the subtraction of them represents the pulse width. Now the leading edge time has been obtained by using the achieved TDC method. So we only need to achieve the trailing edge time measurement in the same TDC channel.

### 2.1 Slice structure of Virtex-5 FPGA

First let us have a review about the structure of the slice resource of Xilinx FPGA. As shown in Fig.2, every slice in Virtex-5 FPGA [15] contains four LUTs (Look-Up Table), four DFFs (D Flip-Flop) and a dedicated fast carry chain which is called CARRY4 Primitive. The carry chain is running upwards and has a height of four bits per slice. For each bit, there is a carry multiplexer (MUXCY) and a dedicated XOR gate as shown in the red circles in Fig.3 and their outputs, COi, Oi, can be stored in DFF which is controlled by a multiplexer.

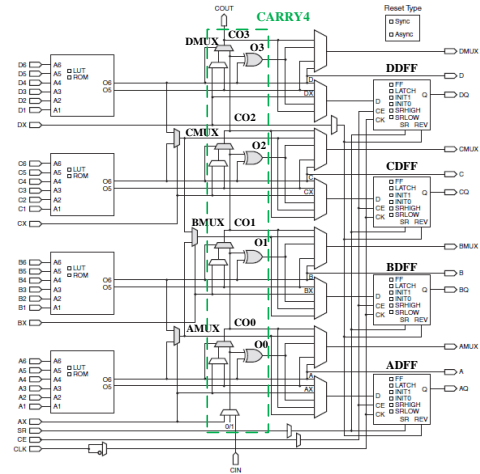

Fig. 2 The structure of slice resource in Xilinx Virtex-5 FPGA

To construct delay cell using the carry chain, all MUXCYs' select inputs must be set '1' to select Cin channels. This makes one input of the XOR gate to be '1'. When a rising edge goes through the carry chain, the outputs, CO0, CO1, CO2 and CO3, of the MUXCYs change to '1' from '0'. When a falling edge comes, the outputs, O0, O1, O2 and O3, of the XOR gates also change to '1' from '0'. So we can use the MUXCY to detect the rising edge and the XOR gate to detect the falling edge.

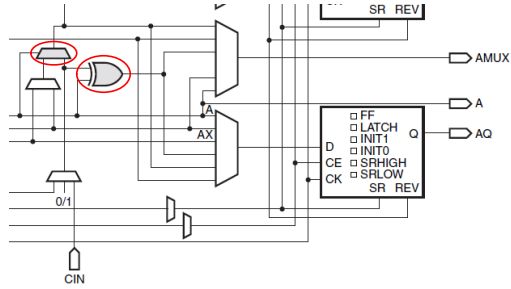

Fig. 3 The enlarged drawing of the first bit of the Slice

**2.2 Implementation**

The block diagram of this method is demonstrated in Fig.4. The 'Hit' signal is feed into the carry chain which is made up of the CARRY4 Primitives. In each CARRY4 Primitive, we take the taps from the port 'CO0' and 'CO3' to construct delay cells for leading edge time measurement and the tap from the port 'O2' for trailing edge time measurement. The taps' status which presents the 'Hit' signal's fine time is stored in the DFFs every clock rising edge. The code is only sent to be encoded in the Encoder and Processing Unite when a true 'Hit' signal comes. Because the detect circuits have the same transitions when a rising edge and a falling edge come, so the encoder unit can be shared by leading edge time measurement and trailing edge time measurement. Finally the fine data and coarse data of the leading edge and trailing edge are written into a same FIFO. Different edges' data are signed by some settled data bits.

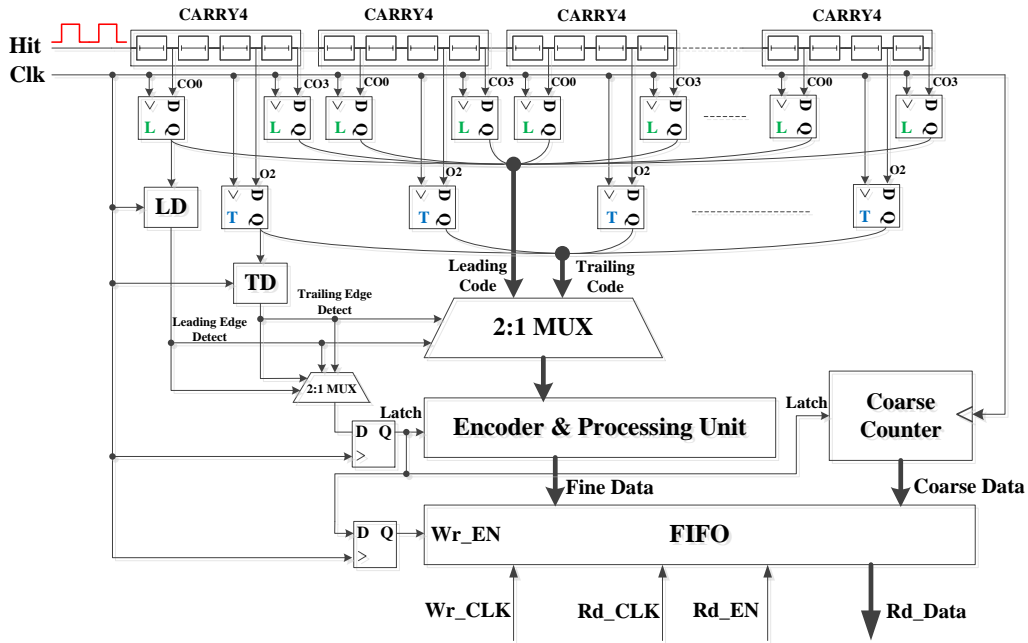

Fig. 4 The architecture of time digitizing system for ETOF upgrade

Two most important features which greatly impact TDC's performance are differential non-linearity (DNL) and integral non-linearity (INL). The larger the DNL and INL are, the worse the TDC's resolution. The non-linearity is mainly caused by the non-uniformity of the dell cells and uneven clock distribution in the registers. For the carry chain resource in Virtex-5 FPGA as shown in Fig.2, there is a multiplexer at the beginning of the CARRY4 Primitive, which is called Asymmetric Factor (AF) in [8]. This makes the four bits' delay time in the CARRY4 Primitive unequal. For making the constructed delay cells as much even as possible to furthest improve the TDC's resolution of time measurement, we take the taps from the port 'CO0' and 'CO3' for leading edge time measurement. The test work and analysis about this content has been talked in [8]. Because the rest two middle bits cannot averagely subdivide one slice, we only take the tap from the port 'O2' for trailing edge time measurement. This means the whole slice or CARRY4 Primitive is as one delay cell to construct the delay chain for trailing edge measurement.

For the encoder unite is shared by leading and trailing edge time measurement, two detect units,

LD (Leading edge Detect) and TD (Trailing edge Detect), are respectively inserted in the first delay cells to indicate a valid rising or falling edge as shown in Fig.4 and we use these two detect unites to decide when and which edge's code should be sent to the encoder unit to be encoded.

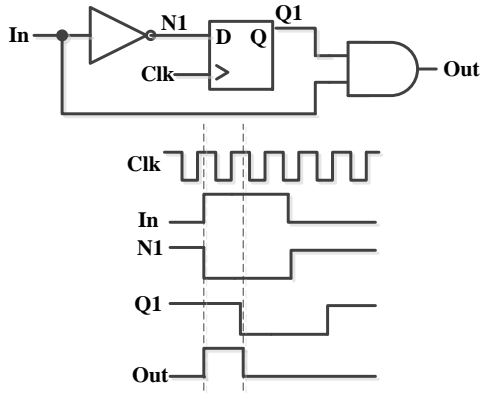

Fig. 5 The specific circuit and timing diagram of hit signal detect unit

On the one hand, their outputs go to control a 2:1 multiplexer to select which edge's code to be encoded. On the other hand, they select themselves through another 2:1 multiplexer to latch the corresponding encoder result after being delayed a clock cycle and enable the FIFO writing operation after two clock cycles' delay. The specific circuit and timing diagram of the detect unit is shown in Fig.5. It applies a NOT gate, a DFF and an AND gate to generate a clock period pulse when a rising edge comes.

**2.3 Dead time**

Since our encoder is based on the dichotomizing search principle, it requires the TDC code to be thermometer code (000…0111…1) and the position of the edge can be determined at the '0-1' border. For guaranteeing the outputs of the registers are thermometer code, the high level and low level of the 'Hit' signal must be respectively one clock cycle at least. Only the signal like this can make the outputs be thermometer code at any case. The limit requires the adjacent two leading edges' smallest time interval must be two clock cycles. This decides that the TDC's dead time of measuring leading edge time is at least two clock cycles, too. For two clock cycles are enough for encoding the first leading edge's code, the second leading edge can be latched into the registers if it comes after two clocks. The encode process doesn't increase the dead time, so the TDC's dead time is two clock cycles in the final.

## 3 Experiments

We designed an evaluation board based on 6U VME standard with a XC5VLX220FF1760 FPGA from Xilinx Virtex-5 family and two channels of the above TDC are implemented on it. The system clock frequency of TDC is 250MHz. The layout of one slice is shown in Fig.6. The blue lines are programmed lines. We can see the bits of CO0, CO3 and O2 are successfully routed and distributed to related delay chains.

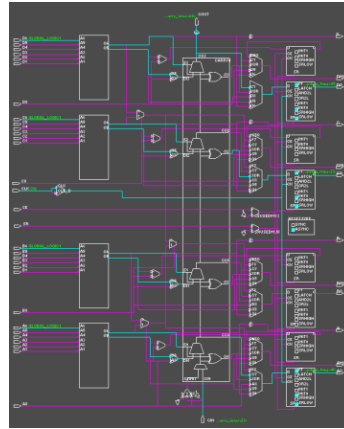

Fig.6. Layout of one slice of the pulse width measurement

Cable delay method [16], [17] is applied to evaluate the TDC's resolution of time measurement. The DNL features of the leading TDC and the trailing TDC are shown in Fig.7 (a) and Fig.8 (a). The bin numbers of them are respectively 105 and 51. The LSB of leading TDC is about 38ps and the LSB of trailing LSB is about 78ps. As we expect, the LSB of the Trailing TDC is almost two times of the Leading TDC. Both the two TDCs' DNL have some large value. This results from the structure of the Xilinx FPGA.

The resolution of the leading edge time and the trailing edge measurement is about 15ps and 21ps as shown in Fig.7 (b) and Fig.8 (b). Although the time resolution of trailing edge time measurement is worse than leading edge time measurement, but it is enough for Time-Walk correction in most high energy physical experiment applications.

One hit signal with constant 23ns pulse width is sent to one channel and the resolution of pulse width measurement is about 37ps which is shown in Fig.9.

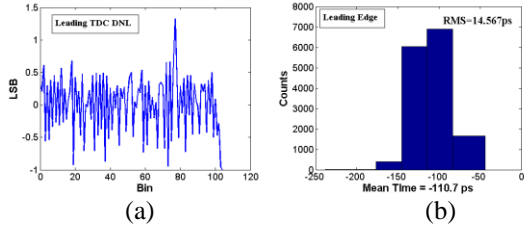

Fig.7. (a) DNL feature of Leading TDC; (b) Time resolution of Leading TDC

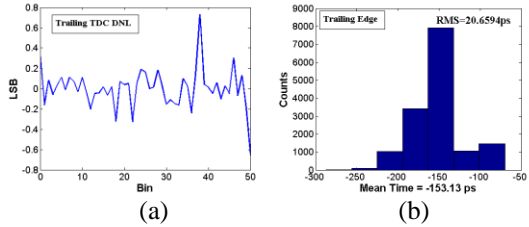

Fig.8. (a) DNL feature of Trailing TDC; (b) Time resolution of Trailing TDC

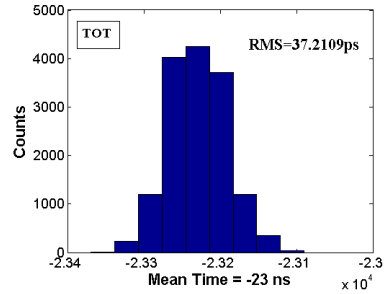

Fig.9. Resolution of pulse width measurement with one TDC channel

## 4 Conclusions

In this paper, we presented a method of measuring pulse width implemented in a Xilinx FPGA. This method implements time measurement and pulse width measurement in one TDC channel. This greatly increases the TDC's integrity and companies with reducing the whole cost. The TDC's performance is (1) leading edge time measurement resolution: ~15ps; (2) trailing edge time measurement resolution: ~21ps; (3) pulse width measurement resolution: ~37ps; (4) dead time: ~10ns.

## References

1. J. Schambach et al. IEEE Nucl. Sci. Sym. Conf. Rec., 2006, 1: 485-488,
2. F. Marcastel et al. ALICE Collaboration, ALICE addendum to the technical design report of the time of flight system (TOF) 2002
3. Huanhuan Fan et al. IEEE Trans. Nucl. Sci., 2013, 60(5): 1-7
4. http://tdc.web.cern.ch/tdc/hptdc/docs/hptdc_manual_ver2.2.pdf, CERN/EPMIC, Mar. 2004
5. Jian Song et al. IEEE Trans. Nucl. Sci., 2006, 53(1): 236-241
6. Jinhong Wang et al. IEEE Trans. Nucl. Sci., 2010, 57(2): 446-450
7. Jinhong Wang et al. IEEE Trans. Nucl. Sci., 2011, 58(4): 2011-2018
8. Lei Zhao et al. IEEE Trans. Nucl. Sci., 60(3): 2013, 3532-3536
9. J. Wu et al. Proc. IEEE NSS, 2003, 1: 177-181
10. Waugh et al. IEEE Trans. Nucl. Sci., 1968, 15(3): 509-517
11. Michael O. B. et al. IEEE Trans. Nucl. Sci., 1979, 26(1): 422-427
12. H. Lim et al. review of scientific instruments, 2003, 74(6): 3115-3119
13. F. Anghinolfi et al. Nucl. Instr. Meth, 2004, A533: 183
14. F. Anghinolfi et al. IEEE Trans. Nucl. Sci., 2004, 51(5): 1974-1978
15. "Virtex-5 FPGA User Guide," UG190 (v5.3), Xilinx Inc. May 17, 2010
16. J. Doernberg et al. IEEE J. Solid-State Circuits, 1984, 19(6): 820-827
17. M. Mota et al. Ph.D. dissertation, Dept. Elect. Eng. Comput., Tech. Univ. Lisbon, Lisbon, Portugal, Oct. 2000